\documentclass[fleqn,twoside]{article}
\usepackage{espcrc2}
\usepackage{epsfig}

\usepackage{graphicx}
\usepackage[figuresright]{rotating}

\newcommand{\AmS}{{\protect\the\textfont2
  A\kern-.1667em\lower.5ex\hbox{M}\kern-.125emS}}

\title{Lepton-nucleus scattering in the impulse approximation regime}

\author{O. Benhar\address[INFN]{INFN, Sezione di Roma. I-00185 Roma, Italy}
\address[DF]{Dipartimento di Fisica, Universit\`a ``La Sapienza". I-00185 Roma, Italy}, %
N. Farina\addressmark[DF], %
H. Nakamura\address[WAS]{Department of Physics, Waseda University. Tokyo 169-8555, Japan}, %
M. Sakuda\address[OKA]{Department of Physics, Okayama University. Okayama, 700-8530, Japan}, %
R. Seki\address[CALST]{Dept. of Physics, California State University. Northridge,
California 91330, USA}\address[KELLOG]{W.K. Kellog Radiation Laboratory, Caltech.
Pasadena, California 91125, USA}
        }

\begin{document}

\begin{abstract}
We discuss theoretical calculations of electron- and neutrino-nucleus scattering, 
carried out using realistic nuclear spectral functions and including the effect
of final state interactions.
Comparison between electron scattering data and the calculated inclusive cross
sections off oxygen shows that the Fermi gas model
fails to provide a satisfactory description of the measured cross sections, and
inclusion of nuclear dynamics is needed. The role of Pali blocking in charged-current 
neutrino induced reactions at low $Q^2$ is also analyzed.
\vspace{1pc}
\end{abstract}

\maketitle

\section{INTRODUCTION}

The field of neutrino physics is rapidly developing after
atmospheric neutrino oscillations and solar neutrino oscillations
have been established \cite{Kajita,Solar,KamLAND}.
Most neutrino experiments measure energy
and angle of muons produced in neutrino-nucleus interactions and
reconstruct the incident neutrino energy, which determines the neutrino
oscillations.
To reduce the systematic uncertainty of these experiments it is vital that
the nuclear response to weak interactions be under control at a quantitative 
level. A number of theoretical approaches aimed at providing accurate
predictions of neutrino-nucleus scattering observables are
discussed in Refs. \cite{NUINT01,NUINT04}

At $E_{\nu}$=3 GeV or less, quasi-elastic scattering and quasi-free
$\Delta$ production are the dominant neutrino-nucleus processes.
However, reactions in this energy regime are associated with a
wide range of momentum transfer, thus involving different aspects of
nuclear structure.

In this short note we discuss results obtained using the many-body theory
of electron-nucleus scattering in the impulse approximation (IA) regime 
(see, e.g., Ref. \cite{vijay_nuint01} and references therein), as well as
its extension to charged-current neutrino-induced reactions. We focus on 
the energy range $0.7-1.2$ GeV, and analyze inclusive
scattering of both electrons and neutrinos off oxygen,
the main target nucleus in SK, K2K and other experiments.

\section{RESULTS}

Within the IA, the cross section of the process
$e+A \rightarrow e^\prime+X$ can be written in the form 
(see, e.g. Ref. \cite{gangofsix})
\begin{equation}
\left( \frac{d\sigma_A}{d\Omega_{e^\prime} d\nu} \right)_{IA} = 
\int d^3 p\ dE P({\bf p},E)\ \frac{d\sigma_N}{d\Omega_{e^\prime} d\nu}\ ,
\label{sigm:1}
\end{equation}
where $\nu = E_e - E_{e^\prime}$ is the electron energy loss, 
the spectral function $P({\bf p},E)$ yields the probability 
of finding a nucleon with momentum ${\bf p}$ and removal energy $E$ 
in the nuclear target and the differential cross section 
$d\sigma_N/d\Omega_{e^\prime} d\nu$ describes 
the elementary electron-nucleon scattering process.

The cross section of Eq. (\ref{sigm:1}) is obtained under the assumption 
that final state interactions (FSI) between the struck nucleon
and the spectator particles be negligible. However, coincidence $(e,e^\prime p)$ data 
unequivocally show that FSI play a significant role, leading to 
a sizable reduction of the outgoing proton flux (see, e.g. Ref. \cite{daniela})

A theoretical approach to include the effect of FSI in inclusive processes 
has been developed in Ref. \cite{gangofsix}. The resulting cross section 
can be written in the convolution form
\begin{equation}
\frac{d\sigma}{d\Omega_{e^\prime} d\nu} = \int d\nu^\prime \
f_{{\bf q}}(\nu - \nu^\prime)\
\left( \frac{d\sigma}{d\Omega_{e^\prime} d\nu^\prime} \right)_{IA} \ .
\label{sigma:FSI}
\end{equation}
The folding function $f_{{\bf q}}(\nu)$ appearing in the above equation, 
that reduces to a $\delta$-function in absence of FSI, is simply related to the 
propagator of the struck nucleon, evaluated within the 
eikonal approximation treating the spectator particles as fixed scattering
centers \cite{gangofsix}.

\begin{figure}[htb]
\vspace{9pt}
{\psfig{figure=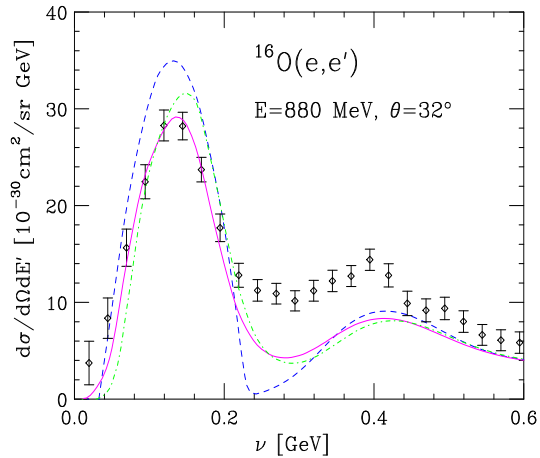,angle=00,width=7.0cm,height=6.0cm}}
\caption{
Cross section of the process $^{16}O(e,e^\prime)$
at beam energy 880 MeV and electron scattering angle 32$^\circ$.
Solid line: full calculation (Eq. (\ref{sigma:FSI})). Dot-dash line: 
IA calculation, carried out neglecting FSI effects (Eq. (\ref{sigm:1})). 
Dashed line: FG model with $p_F = 225$ MeV and $\epsilon = 25$ MeV. The experimental 
data are from Ref.\protect\cite{LNF}.
}
\label{fig:1}
\end{figure}

In Fig. \ref{fig:1} the $(e,e^\prime)$ cross section off oxygen calculated from 
Eqs. (\ref{sigm:1}) and (\ref{sigma:FSI}) using a spectral function obtained within 
the Local Density Approximation \cite{LDA} is compared to the data of Ref. \cite{LNF}.

The theoretical calculation, {\it involving no adjustable parameters}, provides a fairly 
accurate account of the measured cross sections in the region of the quasi-elastic peak. 
The effect of FSI, leading to a shift and a quenching of the peak, is clearly visible.
For reference, the figure also shows the results of the Fermi gas (FG) model, 
corresponding to Fermi momentum $p_F = 225$ MeV and nucleon removal 
energy $\epsilon = 25$ MeV, which appear to largely overestimate the data. 
The failure of the theoretical calculations to reproduce the measured cross section in 
the region of the $\Delta$-production peak is likely to be due to deficiencies in the 
description of the elementary electron-nucleon cross section \cite{PRD}.

In addition to dynamical FSI, arising from by nuclear interactions, statistical 
correlations, leading to Pauli blocking of the phase space available to the
knocked-out particle, must be also taken into account.
A rather crude prescription to estimate the effect of Pauli blocking amounts to
modifying the spectral function through the replacement
\begin{equation}
P({\bf p},E) \rightarrow P({\bf p},E)
\theta(|{\bf p} + {\bf q}| - {\overline p}_F)
\label{pauli}
\end{equation}
where ${\overline p}_F$ is the average nuclear Fermi momentum.

The effect of Pauli blocking is hardly visible in the differential cross section
shown in Fig. \ref{fig:1}, as the kinematical setup corresponds to
$Q^2 > 0.2$ GeV$^2$ at the quasi-elastic peak. 
On the other hand, this effect becomes very large at lower $Q^2$.
 
\begin{figure}[hbt]
\centerline{\psfig{figure=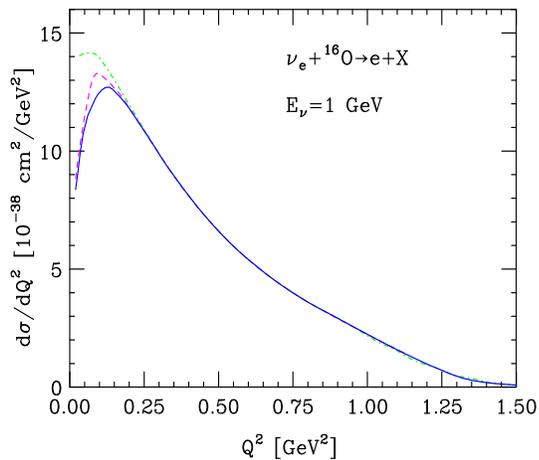
,angle=00,width=7.00cm,height=6.00cm}}
\caption{\small Differential cross section $d\sigma/dQ^2$
for neutrino energy $E_\nu= 1$ GeV. The dot-dash line shows the IA results,
while the solid and dashed lines have been obtained using the modified
spectral function ef Eq. (\protect\ref{pauli}), with and without inclusion
of FSI, respectively.
.}
\label{fig:2}
\end{figure}

This feature is illustrated in Fig. \ref{fig:2}, showing 
the calculated differential cross section $d\sigma/dQ^2$ of the process
$\nu_e~+~^{16}O~\rightarrow~e~+~X$, for neutrino energy $E_\nu= 1$ GeV. 
The dashed and dot-dash lines correspond to the
IA results with and without inclusion of Pauli blocking, respectively. It
clearly appears that the effect of Fermi statistic in suppressing scattering
shows up at $Q^2 < 0.2$ GeV$^2$ and becomes very large at lower $Q^2$. The results of
the full calculation, in which dynamical FSI are also included, are displayed as a full line.
The results of Fig. \ref{fig:2} suggest that Pauli blocking and FSI may explain
the deficit of the measured cross section at low $Q^2$ with respect to the
predictions of Monte Carlo simulations \cite{Ishida}.

\section{SUMMARY}

We have employed an approach based on nuclear many-body theory to compute  
inclusive electron- and neutrino-nucleus scattering cross sections in the 
kinematical region corresponding to beam energy $\sim$ 1 GeV, relevant to many 
neutrino oscillation experiments.

In the region of the quasi-elastic peak, the results of our calculations account
for the measured $^{16}O(e,e^\prime)$ cross sections at beam energies between
700 MeV and 1200 MeV and scattering angle 32$^\circ$ with an accuracy better
than 10 \% \cite{PRD}. It must be emphasized that the ability to yield
quantitative predictions over a wide range of beam energies is critical
to the analysis of neutrino experiments, in which the energy of the incident
neutrino is not known, and must be reconstructed from the kinematics of the
outgoing lepton.

In the region of quasi-free $\Delta$ production theoretical predictions
significantly underestimate the data. In view of the fact that the calculated 
cross sections are in close agreement with the data at higher energies \cite{PRD}, 
i.e. in the region in which inelastic contributions largely dominate,
this problem appears to be mainly ascribable to uncertainties in the
description of the nucleon structure functions at low $Q^2$.

The effect of Pauli blocking, not included in the IA picture, while being hardly 
visible in the lepton energy loss spectra, produces a large effect
on the $Q^2$ distributions at $Q^2 < 0.2$ GeV$^2$, and must therefore be taken
into account.

\end{document}